\documentclass{article}

\usepackage[utf8]{inputenc}
\usepackage[T1]{fontenc}
\usepackage{amsmath, amssymb}
\usepackage[colorlinks=false, linkcolor=black, citecolor=black, urlcolor=black, hidelinks]{hyperref}
\usepackage{geometry}
\usepackage{graphicx}
\usepackage{authblk}
\usepackage[numbers,square]{natbib}
\usepackage{placeins}

\title{\textbf{Modeling the effect of grain boundary diffusivity and trapping on hydrogen transport using a phase-field compatible formulation}\\
{\small Published in: \href{https://doi.org/10.1016/j.ijhydene.2023.11.270}{International Journal of Hydrogen Energy 55 (2024) 1445–1455}}}

\author[1]{Abdelrahman Hussein\thanks{Corresponding author: a.h.a.hussein@outlook.com}}
\author[2]{Byungki Kim}
\author[1]{Tom Depover}
\author[1]{Kim Verbeken}
\affil[1]{Department of Materials, Textiles and Chemical Engineering, Ghent University, Technologiepark 46, B-9052 Ghent, Belgium}
\affil[2]{School of Mechatronics Engineering, Korea University of Technology, Cheonan, Chungnam 31253, Republic of Korea}
\date{} 

\begin{document}

\maketitle

\begin{abstract}
    Hydrogen grain boundary (GB) trapping is widely accepted as the main cause for hydrogen induced intergranular failure. Several studies were conducted to unveil the role of GBs on hydrogen transport; however, a clear understanding is yet to be attained. This is due to the limitations of the state-of-the-art experimental procedures for such highly kinetic processes. In this study, we aim at providing a deeper understanding of hydrogen-GB interactions using full-field representative volume element (RVE). The phase-field method is chosen for generating RVEs, since it is the an appropriate numerical tool to represent GBs. A novel fully-kinetic formulation for hydrogen diffusion and GB trapping is presented, which is compatible with the phase-field based RVEs. GB diffusivity ($D_\mathrm{gb}$) and trap-binding energy ($E_\mathrm{gb}$) were used as parameters to understand the interactions between diffusion and GB trapping. Uptake and permeation simulations were performed with constant and gradient occupancy boundary conditions respectively. In both cases, increasing $E_\mathrm{gb}$, increased the hydrogen GB occupancy. The permeation simulations showed that the hydrogen flux along the GBs increased with increasing both, $D_\mathrm{gb}$ and, surprisingly, $E_\mathrm{gb}$. Since trapping increases the hydrogen occupancy along GBs, it also increases the occupancy gradients, resulting in a higher flux. This led to the conclusion that, in the case of an external occupancy gradient, GB trapping and diffusion cooperate, rather than compete, to increase the hydrogen flux. On the other hand, the decisive factor for the retention of hydrogen at the GBs in permeation simulations was $D_\mathrm{gb}$ rather than $E_\mathrm{gb}$.
\end{abstract}

\section{Introduction}
Hydrogen embrittlement (HE) is one of the long-standing problems for structural materials. The topic is gaining increasing momentum driven by the quest for a sustainable energy carrier. One of the challenges in handling hydrogen is that it can negatively impact the integrity and durability of the infrastructure used in storage and transport, increasing the risk of catastrophic failure \cite{Okonkwo2023, Laureys2022, Cauwels2022, Bouledroua2020}. The complexity of HE analysis lies in the multitude of hydrogen-microstructure interactions. Interface and GB trapping is believed to be the most important factor leading to the hydrogen enhanced decohesion damage mechanism (HEDE), which is the term describing the infamous hydrogen mediated intergranular damage. As such, understanding the equilibrium and kinetics of hydrogen-GB trapping is key to understanding the HEDE mechanism \cite{Bechtle2009}.

Including hydrogen trapping effects to Fick's second law of diffusion \cite{Nagumo2016} was achieved in 1963 by McNabb and Foster \cite{nabb1963}. In 1970, Oriani \cite{Oriani1970} presented the thermodynamic equilibrium between hydrogen in lattice and trapping sites, which is a form of the McLean isotherm. Under the postulate of local equilibrium, i.e. very fast trapping kinetics, and low hydrogen lattice solubility, a simpler equation than that of McNabb and Foster was achieved. The main difference between both models is that the McNabb and Foster model is more general and can represent transient trapping, while Oriani's model can be used for slower time-scale processes compared to trapping. Bhadeshia et al. \cite{Song2013} termed both equations as \emph{kinetic} and \emph{local-equilibrium} schemes respectively. These schemes have been widely used for the analysis of hydrogen permeation and thermal desorption spectroscopy analysis to extract apparent diffusivity, trap-binding energy and trap-density \cite{Vecchi2018, Drexler2021}. In 1989, Sofronis and McMeeking \cite{Sofronis1989} included the effect of mechanical stresses in Oriani's local-equilibrium scheme and used it to analyze the hydrogen diffusion near a blunting crack-tip. This has been the core for all subsequent diffusion-coupled mechanics studies \cite{Olden2012, Barrera2016, MartinezPaneda2018, FernandezSousa2020}.

Full-field mesoscale models that explicitly represent microstructure features are gaining increasing interest in recent years \cite{Rimoli2010, Ilin2014, Charles2017,Yan2017, Hassan2018, Hussein2021} as they support and can provide deeper understanding for experimental results since they can locally resolve hydrogen microstructure interactions, which are difficult to resolve experimentally \cite{Koyama2017}. Regarding hydrogen GB diffusivity and trapping, several studies were reported in the literature \cite{Jothi2015, Ilin2016, Diaz2019,Diaz2020}. These studies were based on RVEs generated primarily by Voronoi tesselation and were solved using the finite element method, which requires tricky meshing strategies especially for GBs and triple junctions. More importantly, the GBs were treated as separate materials, which required different equations for lattice and GB regions. 

A plausible alternative to the geometric approach for RVE generation used in the aforementioned discussion is the physics approach based on the phase-field method \cite{Bargmann2018}. Phase-field based RVEs are particularly convenient for mesoscale modeling of hydrogen-microstructure interactions since explicit microstructure features, like GBs, can be identified from the phase-field order-parameters. The approach is widely used in modeling the effect of segregation on GB dynamics \cite{Groenhagen2007, Kim2008, Abdeljawad2015, Kundin2021}. A static phase-field based RVE, i.e. without considering further evolution of the order-parameters, can be used in modeling hydrogen GB interactions. To the best of our knowledge, Zhang et al. \cite{Zhang2021} presented the only attempt for such approach. However, in their formulation, they did not address the equilibrium properties, and thus, proper connection to the trap-binding energy values. Furthermore, the diffusion equation was not presented, leading to the requirement of unnecessary thermodynamic parameters. This led to non-linear steady-state concentration baseline profile, which we believe is not correct since it contradicts the straight-line baseline profile predicted by Fick's first law of diffusion.

In this work, we address these drawbacks with a focus on the profile of the double-obstacle potential used in the multiphase-field method \cite{Steinbach2009} and present expressions for both equilibrium and kinetic cases. In section \ref{section: modeling}, from an expression of the volumetric free energy functional, we derive an expression for the equilibrium hydrogen distribution as well as a fully kinetic mass transport equation, both in terms of occupancy $(\theta)$. The transformation from the unitless occupancy $\theta$ to concentration $c$ (in units of $\mathrm{mol/m^3}$) in post processing as well as the implication of the phase-field numerical GB thickness are briefly discussed. In section \ref{section: RVE} we apply the model to a 2D polycrystalline RVE and perform uptake and permeation simulations where we investigate the effect of trap-binding energy $\Delta E_\mathrm{gb}$ and GB diffusivity $D_\mathrm{gb}$. The interaction of diffusion and trapping is thoroughly discussed in section \ref{section: results} in terms of full-field and RVE averaged results. 

\section{Modeling GB segregation and trapping in the phase-field framework}
\label{section: modeling}

\subsection{Total free energy functional}

The main variable of the phase-field models is the order parameter vector $\phi_i$, $i=1,\dots, N$ describing the spatial distribution of number of grains $N$. The total free energy functional that describes the total energy of the system including segregation\footnote{We use the terms segregation and trapping interchangeably} effect can be expressed as 

\begin{equation}
  F = \int_V \Bigl[f^\mathrm{ch} + f^\mathrm{trap}_\mathrm{gb} + f^\mathrm{intf} \Bigr] dV
  \label{equation: functional}
\end{equation}

\noindent where $f^\mathrm{ch}$ is the chemical free energy density accounting for the energy of the homogenous system, in this case the lattice, $f^\mathrm{trap}_\mathrm{gb}$ is the segregation -- or trapping -- energy density. The last term contributes only to GB dynamics, i.e. the evolution $\partial \phi / \partial t$, and since we are using static RVE, the effect of segregation on GB dynamics is irrelevant to this study. This term is added only for completeness and has no effect on the diffusion. On the other side, during the generation of the RVE, as discussed in section \ref{section: RVE}, this will be the only term used for normal grain growth. 

Since hydrogen is located at the interstitial sites of the host metal lattice, $f^\mathrm{ch}$ will be expressed using the compound energy formalism (CEF) with a two-sublattice model \cite{Sundman2018}. We use as an example $\alpha-$Fe with a BCC lattice as host metal and hydrogen residing in the tetrahedral interstitial sites \cite{Krom2000} as shown in Fig.(S1). In CEF, the model can be denoted as $\mathrm{(Fe)}_1 \mathrm{(H, Va)}_6$, where Fe fills the BCC sites and H and Va (vacancies) intermix at the interstitial sites. $f^\mathrm{ch}$ can be expressed as

\begin{equation}
  f^\mathrm{ch} = (1 - \theta) G_\mathrm{Fe:Va} + \theta G_\mathrm{Fe:H} + \\ \frac{6RT}{V_\mathrm{m}} \left[(1 - \theta) \ln(1-\theta) + \theta \ln \theta\right]
  \label{Eq: fchem}
\end{equation}

\noindent Where $\theta$ is the occupancy of hydrogen in the interstitial sites. In the CALPHAD community, the term site-fraction and the symbol $y$ are used, as well as in the model of Svoboda and Fischer \cite{Svoboda2012}. However, we will use \emph{occupancy} denoted by $\theta$ following the hydrogen community. $G_\mathrm{Fe:Va}$ and $G_\mathrm{Fe:H}$ are the Gibbs energy of the end members. $R$ is the universal gas constant, $T$ is the absolute temperature and $V_\mathrm{m}$ is the molar volume of the host lattice. The last term in Eq.(\ref{Eq: fchem}) is the entropy due to the intermixing of Va and H in the interstitial sites. Since Fe totally fills the BCC sites, it has no contribution to the entropy of mixing. 

While there are various models in the literature to describe the various aspects of  $f^\mathrm{trap}_\mathrm{gb}$ \cite{Abdeljawad2015}, we use a simplified version of the model proposed by Kim and Park \cite{Kim2008} due to its compatibility with the multiphase-field method. $f^\mathrm{trap}_\mathrm{gb}$ is constructed to decrease, inducing GB segregation, with increasing solute as 

\begin{equation}
  f^\mathrm{trap}_\mathrm{gb} = g(\phi)(1 - \kappa_\mathrm{gb} \theta)
\end{equation}

\noindent Where $g(\phi) = \sum_{\alpha=1}\sum_{\beta>\alpha} \phi_\alpha \phi_\beta$ is the double-obstacle potential, which is the function representing the GBs. This choice ensures that $g(\phi) \neq 0$ at the GBs, while $g(\phi) = 0$ otherwise. $\kappa_\mathrm{gb}$ is a parameter to set the ratio between the lattice and GB occupancy, and will be related to the trap-binding energy as discussed in the next section. 

\subsection{Equilibrium and GB segregation isotherm}
\label{section: Equilibrium}

\begin{figure}[hbt!]
  \includegraphics[width=8cm]{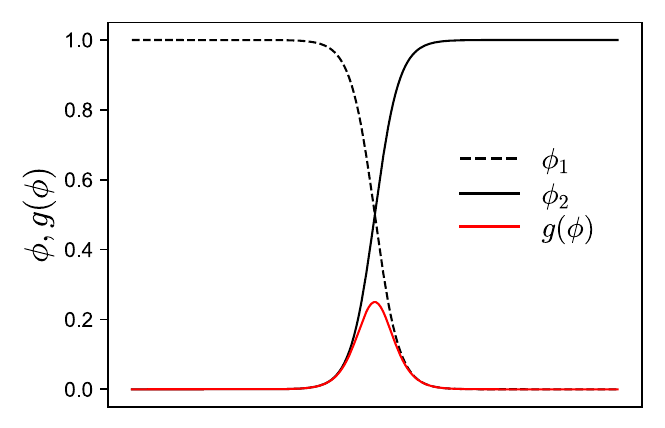}
  \centering
  \caption{A cross-section of along the infinite bicrystal interface showing the profile of $\phi_1$, $\phi_2$ and the interface function $g(\phi)$}
  \label{fig: Psi}
\end{figure}

In this section, we derive the equilibrium occupancy distribution between GB center and the lattice bulk far away from the GB. First, to facilitate the analysis let us consider an infinite bicrystal with flat interface, where the two grains are represented by $\phi_1$ and $\phi_2 = 1-\phi_1$, and thus $g(\phi) = \phi_1 \phi_2$ as shown in Fig.(\ref{fig: Psi}). 

At equilibium, the chemical potential in the lattice $\mu_\mathrm{L}$ far away from the GB, i.e. $g(\phi) = 0$ and occupancy $\theta_\mathrm{L}$ is expressed by 

\begin{equation}
  \mu_\mathrm{L} = \frac{\partial F}{\partial \theta} |_{\theta_\mathrm{L}} = G_\mathrm{Fe:H} - G_\mathrm{Fe:Va} + \frac{6RT}{V_\mathrm{m}} \ln \Bigl(\frac{\theta_\mathrm{L}}{1-\theta_\mathrm{L}} \Bigr)
\end{equation}

The chemical potential at the center of the GB with $g(\phi) = 1/4$ and occupancy $\theta^\mathrm{center}_\mathrm{gb}$ can be expressed as

\begin{equation}
  \mu_\mathrm{gb} = \frac{\partial F}{\partial\theta} |_{\theta^\mathrm{center}_\mathrm{gb}} = G_\mathrm{Fe:H} - G_\mathrm{Fe:Va} \\ + \frac{6RT}{V_\mathrm{m}} \ln \Bigl(\frac{\theta^\mathrm{center}_\mathrm{gb}}{1-\theta^\mathrm{center}_\mathrm{gb}} \Bigr) - \frac{\kappa_\mathrm{gb}}{4}
\end{equation}

The equilibrium occupancy at $\theta^\mathrm{center}_\mathrm{gb}$ can be obtained by constructing a parallel tangent to that at the lattice for a given lattice occupancy $\theta_\mathrm{L}$ \cite{Lejcek2010} as shown schematically in Fig.(S2). Therefore, $\mu_\mathrm{L} = \mu_\mathrm{gb}$ and thus

\begin{equation}
  \begin{split}
    \frac{\theta^\mathrm{center}_\mathrm{gb}}{1-\theta^\mathrm{center}_\mathrm{gb}} &= \frac{\theta_\mathrm{L}}{1-\theta_\mathrm{L}} \exp\Bigl(\frac{\kappa_\mathrm{gb} V_\mathrm{m}}{24RT}\Bigr) \\
    &=\frac{\theta_\mathrm{L}}{1-\theta_\mathrm{L}} \exp\Bigl(\frac{\Delta E_\mathrm{gb}}{RT}\Bigr)
  \end{split} 
  \label{Eq: Equilibrium1}
\end{equation}

\noindent Where $\Delta E_\mathrm{gb}$ is the GB segregation enthalpy or trap-binding energy. As described in \cite{Svoboda2015}, an equilibrium equation in the form of Eq.(\ref{Eq: Equilibrium1}) was first formulated by McLean \cite{McLean1957} in 1957, famously known in the segregation community as the McLean-Langmuir isotherm \cite{Lejcek2010}. It was first introduced to the hydrogen community by Oriani \cite{Oriani1970} in 1970 with the dilute approximation $\theta_\mathrm{L}\ll1$. This equilibrium construction will naturally result in the \emph{local-equilibrium} condition hypothesized by Oriani, and will be further elaborated in the permeation simulations in section \ref{section: PermeationDiffusivity}. The parameter $\kappa_\mathrm{gb}$ is finally related to $\Delta E_\mathrm{gb}$ as

\begin{equation}
  \kappa_\mathrm{gb} = -\frac{24 \Delta E_\mathrm{gb}}{V_\mathrm{m}}
  \label{Eq: kappa}
\end{equation}

\noindent The value of $\theta^\mathrm{center}_\mathrm{gb}$ as a function of $\theta_\mathrm{L}$ for different values of $\Delta E_\mathrm{gb}$ according to Eq.(\ref{Eq: Equilibrium1}) is shown in Fig.(S3). Using Eq.(\ref{Eq: kappa}) for any RVE with GBs represented by $g(\phi)$ and equilibrium lattice occupancy field $\theta^\mathrm{eq}_\mathrm{L}$, the equilibrium occupancy field $\theta^\mathrm{eq}$ could be obtained from  

\begin{equation}
    \frac{\theta^\mathrm{eq}}{1-\theta^\mathrm{eq}} = \frac{\theta^\mathrm{eq}_\mathrm{L}}{1-\theta^\mathrm{eq}_\mathrm{L}} \exp\Bigl(\frac{\kappa_\mathrm{gb} V_\mathrm{m}g(\phi)}{6RT}\Bigr)
  \label{Eq: Equilibrium}
\end{equation}

\subsection{The diffusion equation}

The diffusional flux of the occupancy field $\theta$ is given by linear irreversible thermodynamics 

\begin{equation}
  \vec{J} = -\theta M \nabla \Bigl( \frac{\delta F}{\delta \theta} \Bigr) 
\end{equation}

\noindent where $M$ is the diffusional mobility. We have 

\begin{equation}
  \nabla \frac{\delta F}{\delta \theta} = \nabla \frac{\partial F}{\partial \theta} = \frac{\partial F^2}{\partial \theta^2} \nabla \theta + \frac{\partial F^2}{\partial \theta \partial g(\phi)} \nabla g(\phi)
\end{equation}

\noindent In the dilute limit of the lattice, the diffusivity is related to the diffusional mobility by $M=D/RT$

\begin{equation}
  \begin{split}
    \vec{J} &= - D \Bigl( \frac{6}{V_\mathrm{m}} \nabla \theta - \frac{\kappa_\mathrm{gb} \theta}{RT} \nabla g(\phi) \Bigr) \\
  &= \vec{J}_\mathrm{\theta} + \vec{J}_\mathrm{gb}
  \end{split}
  \label{Eq: flux}
\end{equation}

\noindent In the case where the GB diffusivity $D_\mathrm{gb}$ is different from that in the lattice $D_\mathrm{L}$, we adopt the expression proposed in \cite{Groenhagen2007} for the diffusivity field\footnote{Note that this is a field expression since $D = D(\phi)$ and $\phi=\phi(x)$. Please refer to section \ref{section: PermeationDiffusivity} and Fig.(\ref{fig: Diffusivity}) for further discussions.} $D$

\begin{equation}
  D = D_\mathrm{L} \Bigl( \frac{D_\mathrm{gb}}{D_\mathrm{L}} \Bigr)^{4g(\phi)}
  \label{Eq: Diffusivity}
\end{equation}

\noindent The time evolution of the occupancy is governed by the conservation equation 

\begin{equation}
  \frac{\partial \theta}{\partial t} = -\nabla \cdot \vec{J} 
  \label{Eq: diffusion}
\end{equation}

\begin{figure}[hbt!]
  \includegraphics[width=8cm]{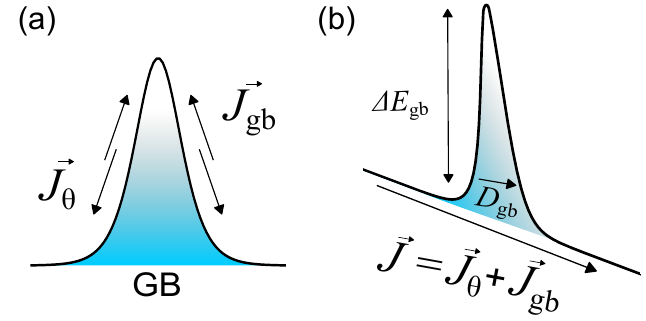}
  \centering
  \caption{Schematic showing (a) the direction of interaction of the diffusion $\vec{J}_\mathrm{\theta}$ and trapping $\vec{J}_\mathrm{gb}$ flux across
   the GB and (b) the resultant flux $\vec{J}$ in the presence of external concentration gradient. The height of the GB potential is determined by $\Delta E_\mathrm{gb}$, while $D_\mathrm{gb}$ determines the flux magnitude of trapping and diffusion across GBs.}
  \label{fig: Psi_schematic}
\end{figure}

\noindent There are a few remarks to make on Eq.(\ref{Eq: flux}):

\begin{itemize}

  \item Following Bhadeshia's terminology \cite{Song2013} discussed in the introduction, the equation is a \emph{fully kinetic} model describing two kinetic processes as shown schematically in Fig(\ref{fig: Psi_schematic}a): a conventional Fickian diffusion process with flux $\vec{J}_\mathrm{\theta}$ in the direction opposite to the occupancy (concentration) gradient, and a trap-filling process driven by the GB potential $g(\phi)$ with a flux $\vec{J}_\mathrm{gb}$ directed towards the GB center, i.e. in the direction of $\nabla g(\phi)$. This is in agreement with the description of Koyama et al. \cite{Koyama2017}. Therefore, throughout this article we use the term \emph{diffusion} for occupancy gradient driven flux and \emph{trapping} for $g(\phi)$ (GB) gradient driven flux. The term \emph{permeation} is used for permeation simulations with occupancy gradient boundary conditions as discussed later in section \ref{section: RVE}.
  
  \item The height of the GB potential $g(\phi)$, shown schematically in Fig(\ref{fig: Psi_schematic}b), is determined by the trap-binding energy $\Delta E_\mathrm{gb}$ (implicitly in $\kappa_\mathrm{gb}$), which represents the capacity for hydrogen enrichment. The GB diffusivity $D_\mathrm{gb}$ determines the flux magnitude of both trap-filling and diffusion across the GB.
   
  \item Unlike the center of the GBs where $g(\phi)=1/4$, the center of the triple junctions has a value of $g(\phi)=1/3$, which implicitly accounts for their effect on trapping and diffusivity.
    
  \item Both Eq.(\ref{Eq: Equilibrium}) and (\ref{Eq: diffusion}) solve for the total hydrogen without distinguishing between lattice and trapped hydrogen. Such partitioning can be obtained during post-processing by using the interface function where $g(\phi) \neq 0$ in the GBs and $g(\phi) = 0$ in the lattice. 

  \item Finally, unlike other models in the literature that use concentration, in units of mole or atoms per unit volume,  as the independent variable, here we use the unitless occupancy. The advantage is that the occupancy can be directly related to the thermodynamic equilibrium as in Eq.(\ref{Eq: Equilibrium}). Svoboda and Fischer \cite{Svoboda2012} presented a similar concept and termed it \emph{dimension-free formulation}. They made the conversion to the volumetric molar concentration $c$ using the relation $c=\theta/V$, where $V$ is the molar volume. However, this conversion requires the distinction between the lattice and GB molar volumes $V_\mathrm{m}$ and $V_\mathrm{gb}$ respectively. The latter is an extra parameter that need to be provided, which is not always straightforward to evaluate \cite{Diaz2020, Svoboda2014}. This is an advantage for Eq.(\ref{Eq: flux}), where $V_\mathrm{gb}$ is not mandatory, but rather an optional parameter if $c$ is of interest instead of $\theta$. 

\end{itemize}

Before we proceed to polycrystalline RVE, we investigate the behavior of Eqs.(\ref{Eq: Equilibrium} and \ref{Eq: diffusion}) in a single GB. 

\subsection{Application to a single GB of infinite flat bicrystal}
\label{section: Application}

We use the bicrystal example shown in Fig.(\ref{fig: Psi}) with numerical interface width $\eta=5 \ \mu$m and parameters representing $\alpha$-Fe as host metal listed in Table(\ref{Table: 1D}). Eq.(\ref{Eq: diffusion}) is solved using the finite-difference method with the forward Euler time integration scheme. We used 100 grid points with grid spacing $\Delta x=0.1666 \ \mu$m and a numerical increment adjusted according to the time stability criterion $\Delta t \le \Delta x^2V_\mathrm{m}/12D$. It should be noted that $\Delta t$ has units of $\mathrm{sm^{3}/mol}$. The physical time in units of s is $6\Delta t/V_\mathrm{m}$. The initial occupancy $\theta_0$ was uniform and constant boundary conditions were applied at both ends with $\theta_0=\theta_\mathrm{boundary}=10^{-9}$. 

\begin{table}[hbt!]
  \centering
  \caption{Parameters for 1D bicrystal simulation}
  \begin{tabular}{ll}

  \hline
  Parameter                       &Value                       \\ 
  \hline
  $V_\mathrm{m}$ $\mathrm{(m^3/mol)}$            & $7.09\times10^{-6}$  \\
  $D$   $\mathrm{(m^2/s)}$                       & $1\times10^{-8}$     \\
  $T$   $\mathrm{(K)}$                           & $300$                \\
  $R$   $\mathrm{(J/mol K)}$                     & $8.31$               \\
  $\Delta E_\mathrm{gb}$   $\mathrm{(J/mol)}$   & $-1000$              \\
  \hline               
  \end{tabular}
  \label{Table: 1D}
\end{table}

\begin{figure}[hbt!]
  \includegraphics[width=8cm]{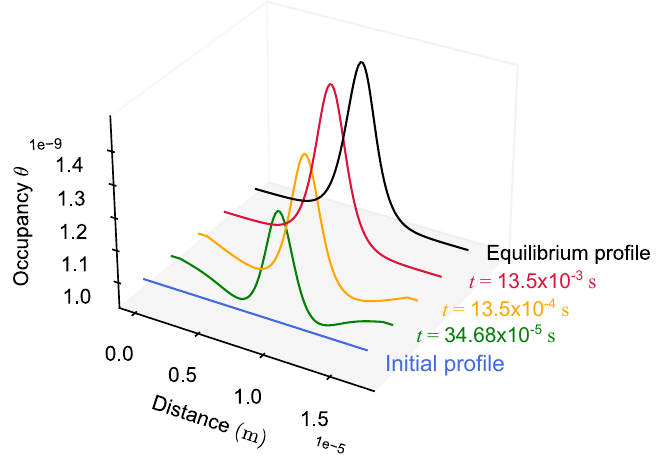}
  \centering
  \caption{The evolution of the occupancy according to Eq.(\ref{Eq: diffusion}) for a single GB with $\Delta E_\mathrm{gb}=-1 \ \mathrm{kJ/mol}$ and the equilibrium profile accoding to Eq.(\ref{Eq: Equilibrium}).}
  \label{fig: Profile}
\end{figure}

Starting from a uniform initial profile, the evolution of trap filling could be seen at various time steps in Fig.(\ref{fig: Profile}). The equilibrium was reached after $13.5 \times10^{-3}$ s. However, it should be noted that in this example we used a relatively small value for the trap-binding energy $\Delta E_\mathrm{gb}$. Larger values for $\Delta E_\mathrm{gb}$ will lead to longer time to reach equilibrium as be further analyzed in the 2D polycrystalline RVE in section \ref{Section: Uptake}. The equilibrium profile using Eq.(\ref{Eq: Equilibrium}) is an exact match with that of Eq.(\ref{Eq: diffusion}). Therefore, Eq.(\ref{Eq: Equilibrium}) can be used when the trap-filling kinetics are not required, i.e, at equilibrium and/or when the processes of interest have slower kinetics.

\subsection{Numerical vs real interface thickness} 
\label{section: Numerical interface}

One of the challenges of the phase-field method is that the real GB thickness can not be resolved in RVEs representing relatively large grains, due to numerical restrictions. In the previous example we used a numerical interface $\eta=5 \mu$m, while real interface width $\eta_\mathrm{real}$ is in the order of 0.5nm, that is four orders of magnitude less. Although this will not have an impact on the maximum value, which is at the center of the GB, it will affect the area under the curve as shown in Fig(S4), and thus, the overall amount of hydrogen in the GB. Therefore, Eq.(\ref{Eq: Equilibrium}) and (\ref{Eq: diffusion}) overestimate total amount of hydrogen, and thus, should be used for qualitative analysis of hydrogen-GB interaction. A method to overcome this limitation is being currently investigated by the authors.

\section{Simulations using 2D polycrystalline RVE}
\label{section: RVE}

In this section we simulate hydrogen diffusion and GB trapping in a 2D polycrystalline RVE using Eq.(\ref{Eq: diffusion}). First, the RVE was generated using the phase-field method for normal grain-growth using the free energy functional

\begin{equation}
  F = \int_V \Bigl[\sum_{\alpha=1}\sum_{\beta>\alpha} \frac{4 \sigma_{\alpha \beta}}{\eta} \Bigl(-\frac{\eta^2}{\pi^2} \nabla\phi_\alpha \cdot \nabla\phi_\beta \\ + \phi_\alpha \phi_\beta \Bigr) \Bigr] dV
  \label{equation: Functional2}
\end{equation}

\noindent Where $\sigma_{\alpha \beta}$ is the interface energy between phases $\alpha$ and $\beta$. The evolution of $\phi$, i.e. microstructure, is given by 

\begin{equation}
  \dot{\phi} = \frac{\pi^2}{8 \eta} \sum^N_{\alpha \neq \beta} \frac{\mu_{\alpha \beta}}{N} \left( \frac{\delta F}{\delta \phi_\alpha} - \frac{\delta F}{\delta \phi_\beta} \right)
\label{Eq:Evolution}
\end{equation}

\noindent Where $\mu_{\alpha \beta}$ is the interface mobility between phases $\alpha$ and $\beta$. We used an isotropic interface energy of 0.5 $\mathrm{Jm^{-2}}$ and interface mobility of $10^{-14} \ \mathrm{m^{2}J^{-1}s^{-1}}$. The resulting RVE with the interface function $g(\phi)$ representing GBs is shown in Fig.(S5a) and the grain size distribution Fig.(S5b). As in the 1D simulations, we used numerical interface width $\eta=5 \ \mu$m.

We performed two sets of simulations: hydrogen uptake and hydrogen permeation. For the uptake simulations, we investigate the effect of trap-binding energy $\Delta E_\mathrm{gb}$ on the amount of hydrogen in the RVE and time to reach equilibrium. In these simulations we used constant boundary conditions (BCs) as shown schematically in Fig.(S6a). The initial occupancy $\theta_0=0$ and $\theta_\mathrm{boundary}=4.08\times10^{-9}$ which is equivalent to lattice concentration $c_\mathrm{L} = 3.453\times10^{-3} \ \mathrm{mol/m^3} = 2.08\times10^{21}\mathrm{ \ atoms/m^3}$ as used by Sofronis and McMeeking \cite{Sofronis1989}. The permeation simulations investigate the effect of GB diffusivity $D_\mathrm{gb}$ and trapping $\Delta E_\mathrm{gb}$ with BCs shown in Fig.(S6b)

The simulations were performed using standard finite-difference method with the explicit Euler method and a 21-point central difference stencil, which has eighth-order accuracy. The grid points were 400 $\times$ 400 or 800 $\times$ 800 with grid spacing $\Delta x = 0.25 \ \mu$m and $\Delta x = 0.125 \ \mu$m respectively. The numerical increment was $\Delta t \le \Delta x^2V_\mathrm{m}/12D_\mathrm{max}$, where $D_\mathrm{max}$ is the maximum diffusivity in the RVE. 

\section{Simulation results and discussion}
\label{section: results}

\subsection{Effect of trap-binding energy on hydrogen uptake}
\label{Section: Uptake}

\begin{figure*}[hbt!]
  \includegraphics[width=15cm]{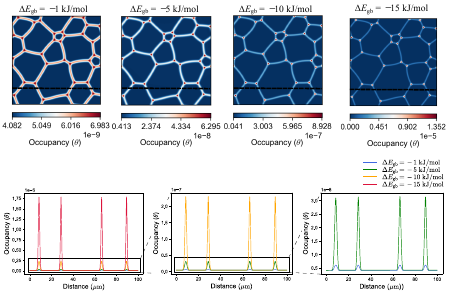}
  \centering
  \caption{The occupancy fields after reaching equilibrium for different GB trap-binding energies $\Delta E_\mathrm{gb}$. In the second row, the occupancy profiles across the GBs along the black dashed lines.}
  \label{fig: UptakeOccupancy}
\end{figure*}

The occupancy field after reaching equilibrium is shown in Fig.(\ref{fig: UptakeOccupancy}) upper row. As expected, increasing $\Delta E_\mathrm{gb}$ lead to orders of magnitude increase in the occupancy at the GBs according to Eq.(\ref{Eq: Equilibrium1}). The profile across the GBs along the dashed lines is shown in the lower row of Fig.(\ref{fig: UptakeOccupancy}). It could be seen that the baselines of the profiles are straight, unlike the non-linear baselines reported in \cite{Zhang2021}, and have a smooth transition between lattice and GB regions compared to the small discontinuity in \cite{Diaz2020}. The baseline occupancy value is the lattice value $\theta_\mathrm{L}$ and is equal to the boundary occupancy $\theta_\mathrm{boundary}$. The peak values are at the GB center and are related to the lattice value according to Eq.(\ref{Eq: Equilibrium1}). Similar to the  discussion in section \ref{section: Application}, an exact equilibrium occupancy field can be obtained using Eq.(\ref{Eq: Equilibrium}).

At the triple junctions, the occupancy values are larger than the equilibrium values predicted by Eq.(\ref{Eq: Equilibrium1}). As discussed before, this is because at the triple junctions $g(\phi)=1/3$, and substituting in Eq.(\ref{Eq: Equilibrium}) will lead to larger values compared to the center of GB where $g(\phi)=1/4$. Due to their more open structure compared to GBs, triple junctions were reported to have larger segregation enthalpy \cite{Stender2011}. This increased hydrogen enrichment at the triple junctions was reported in molecular dynamics simulation \cite{Zhou2021} and was accounted for by assigning higher trap-binding energy to the triple junctions. Translating this to the mesoscale RVEs based on geometric methods \cite{Jothi2015a} required extra processing to identify the triple junctions. In the current case of phase-field based RVEs, this is a desirable side effect where the triple junctions are implicitly accounted for without the need to assign higher trap-binding energy. It should be also noted that quadruple and higher order junctions will be also naturally accounted for using the phase-field based RVEs. Due to their topology, these junctions are more likely to be present in 3D compared to the 2D case used in the current study. 

\begin{figure*}[hbt!]
  \includegraphics[width=15cm]{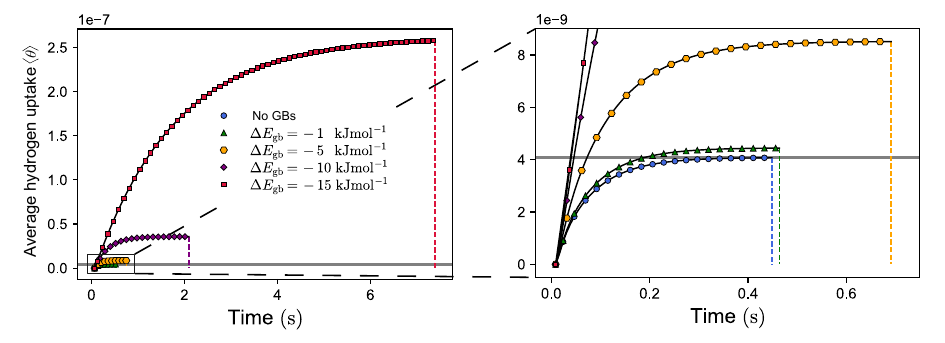}
  \centering
  \caption{Average hydrogen uptake for different GB trap-binding energies $\Delta E_\mathrm{gb}$. The horizontal gray line represents the boundary occupancy $\theta_\mathrm{boundary}$. The vertical dashed lines are to guide the eye for the time to reach equilibrium.}
  \label{fig: UptakeTotal}
\end{figure*}

The average hydrogen uptake $\langle \theta \rangle = 1/A \int \theta dA$ all over the RVE with area $A$ vs time is shown in Fig.(\ref{fig: UptakeTotal}) compared to an RVE with no GBs. The calculations were performed until equilibrium was reached according to the condition $\Delta \theta / \Delta t \le 10^{-4}$. It could be seen that the trap-binding energy affects both the total hydrogen uptake and the time to reach equilibrium, i.e. trap-filling. Higher trapping energy lead to larger uptake as well as longer time to reach equilibrium. Further understanding of the effect of $\Delta E_\mathrm{gb}$ on the interaction between GB trapping and diffusion as well as trap-filling mechanism will be discussed in section \ref{section: PermeationEnergy} using permeation simulations.

\subsection{Effect of grain boundary diffusivity on hydrogen permeation}
\label{section: PermeationDiffusivity}

\begin{figure}[hbt!]
  \includegraphics[width=8cm]{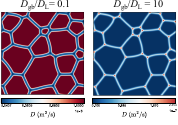}
  \centering
  \caption{Two cases of the diffusivity field $D$ where $D_\mathrm{gb}<D_\mathrm{L}$ and $D_\mathrm{gb}>D_\mathrm{L}$.}
  \label{fig: Diffusivity}
\end{figure}

In this section we analyze the role of GB diffusivity $D_\mathrm{gb}$ from the permeation simulations. Since in the literature a wide range of GB to lattice diffusivity ratios $D_\mathrm{gb}/D_\mathrm{L}$ is reported, which depends, among other factors, on the GB misorientation and trap-binding energy \cite{Zhou2021, Oudriss2012}, we use the ratios 0.01, 0.1, 0.5, 1, 2 and 10 to cover a reasonable range between $D_\mathrm{gb}<D_\mathrm{L}$ and $D_\mathrm{gb}>D_\mathrm{L}$. In these simulations, we used a fixed value of $\Delta E_\mathrm{gb}$ at $\mathrm{-5 \ kJ/mol}$. The diffusivity fields according to Eq.\ref{Eq: Diffusivity} representing the two cases are shown in Fig.\ref{fig: Diffusivity}. It could be seen that the triple junctions diffusivity is larger than GB diffusivity in the case of $D_\mathrm{gb}>D_\mathrm{L}$, while it is smaller in the case of $D_\mathrm{gb}<D_\mathrm{L}$. Therefore, the construction of Eq.\ref{Eq: Diffusivity} results in the amplification of the $D_\mathrm{gb}/D_\mathrm{L}$ ratio at the triple junctions.

\begin{figure*}[hbt!]
  \includegraphics[width=16cm]{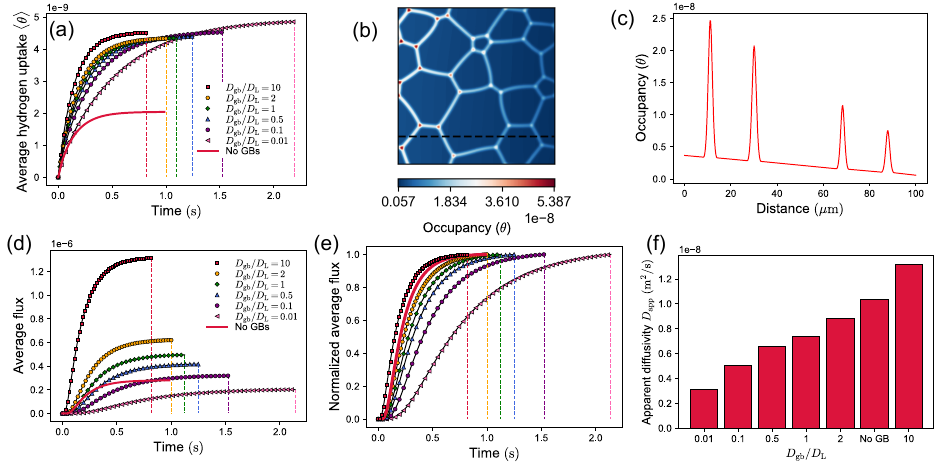}
  \centering
  \caption{(a) The evolution of the average occupancy $\langle\theta\rangle$ over the whole RVE. (b) The equilibrium occupancy field for the case $D_\mathrm{gb}/D_\mathrm{L} = 1$ , which is similar for all other ratios (c) the profile along the dashed line in (b). (d) The average flux at the exit side and (e) the normalized flux. (d) The apparent diffusivity $D_\mathrm{app}$ as a function of $D_\mathrm{gb}/D_\mathrm{L}$.}
  \label{fig: PermeationAverage}
\end{figure*}

The evolution of the average occupancy $\langle \theta \rangle$ until steady-state equilibrium for different ratios $D_\mathrm{gb}/D_\mathrm{L}$ compared to the case of No GBs is shown in Fig.(\ref{fig: PermeationAverage}a). It could be seen that the equilibrium $\langle \theta \rangle$ is almost equal for all cases and larger than the No GBs case. This means that the grain boundary diffusivity has a negligible effect on the hydrogen uptake. Indeed, the equilibrium occupancy fields are only slightly different for all the ratios $D_\mathrm{gb}/D_\mathrm{L}$. The occupancy field for the case $D_\mathrm{gb}/D_\mathrm{L}=1$ is shown in Fig.(\ref{fig: PermeationAverage}b) with the profile along the dashed line shown in Fig.(\ref{fig: PermeationAverage}c). Again, the baseline is a straight line with smooth transition between the lattice and the GBs. The baseline is identical to the lattice occupancy in the case of No GBs. Furthermore, the height of the peaks is higher at the entrance side compared to the exit side. This is because the equilibrium peak occupancy is related to the \emph{local} baseline --or lattice-- occupancy following Eq.(\ref{Eq: Equilibrium1}). This is a clear manifestation of Oriani's local equilibrium hypothesis, which results from the parallel tangent construction of equilibrium as discussed in section \ref{section: Equilibrium}.

The evolution of the average $x-$component of the flux evaluated according to Eq.(\ref{Eq: flux}) at the exit side is shown in Fig.(\ref{fig: PermeationAverage}d). As expected, increasing $D_\mathrm{gb}$ increases the average flux in addition to accelerating the time to reach the steady state equilibrium, as shown by the normalized average flux in Fig.(\ref{fig: PermeationAverage}e). The apparent diffusivity is calculated using the permeation curves according to 

\begin{equation}
  D_\mathrm{app} = \frac{L^2}{6 t_{0.63}}
\end{equation}

\noindent Where $L$ is the length of the RVE and $t_{0.63}$ is the time to reach 63\% of the steady state flux. As expected, $D_\mathrm{app}$ increases with $D_\mathrm{gb}$ as shown in Fig.(\ref{fig: PermeationAverage}f). However, in order to put these numbers into perspective, the case of No GBs should be taken into account. The equilibrium flux for No GBs is comparable to the case $D_\mathrm{gb}/D_\mathrm{L} = 0.1$. On the other side, reaching the steady state in the case of No GBs is only preceded by $D_\mathrm{gb}/D_\mathrm{L} = 10$. While the values of $D_\mathrm{app}$ qualitatively reflect the speed of reaching steady-state, it does not reflect the equilibrium flux. This raises questions on the validity of using $D_\mathrm{app}$ as a measure for studying trapping. A distinction should be made between the speed of reaching steady-state represented by $D_\mathrm{app}$, which is just a representation of $t_{0.63}$ and hydrogen transport speed, which is represented by the flux.

It was experimentally shown that polycrystals have higher steady-state flux compared to single crystals \cite{Oudriss2012, Brass1996, DiazRodriguez2022} as well as shorter time to reach equilibrium. If we consider the No GBs case as single crystal, this trend is in agreement with our results. Similar trend was also reported for full-field model \cite{Diaz2019}. This generally was explained in terms of higher diffusivity for high-angle GBs in Ni \cite{Oudriss2012}. In their work, the terminology \emph{trapping} was used as opposed to \emph{diffusion}. Koyama et al.\cite{Koyama2017} used in-situ silver decoration with permeation test in pure Fe. They showed that the silver covering at the GBs is due to high hydrogen flux. Since it was experimentally shown that the bulk diffusivity of hydrogen is similar for single and polycrystalline pure Fe \cite{Hagi1979}, i.e. $D_\mathrm{L} \approx D_\mathrm{gb}$, they concluded that the high flux at the GBs is due to trapping. Furthermore, they also showed that high angle GBs have higher flux compared to the low angle GBs, due to their larger $\Delta E_\mathrm{gb}$. These experimental observations on the effect of trapping on hydrogen flux at the GBs is in very good agreement with our simulations. As will be discussed next, GB trapping can indeed result in high flux paths. Therefore, we believe a more accurate terminology for the role of GBs is whether they are \emph{high} or \emph{low flux paths} based on the interaction of trapping and diffusion, represented by $\Delta E_\mathrm{gb}$ and $D_\mathrm{gb}$ respectively.

\begin{figure*}[hbt!]
  \includegraphics[width=13cm]{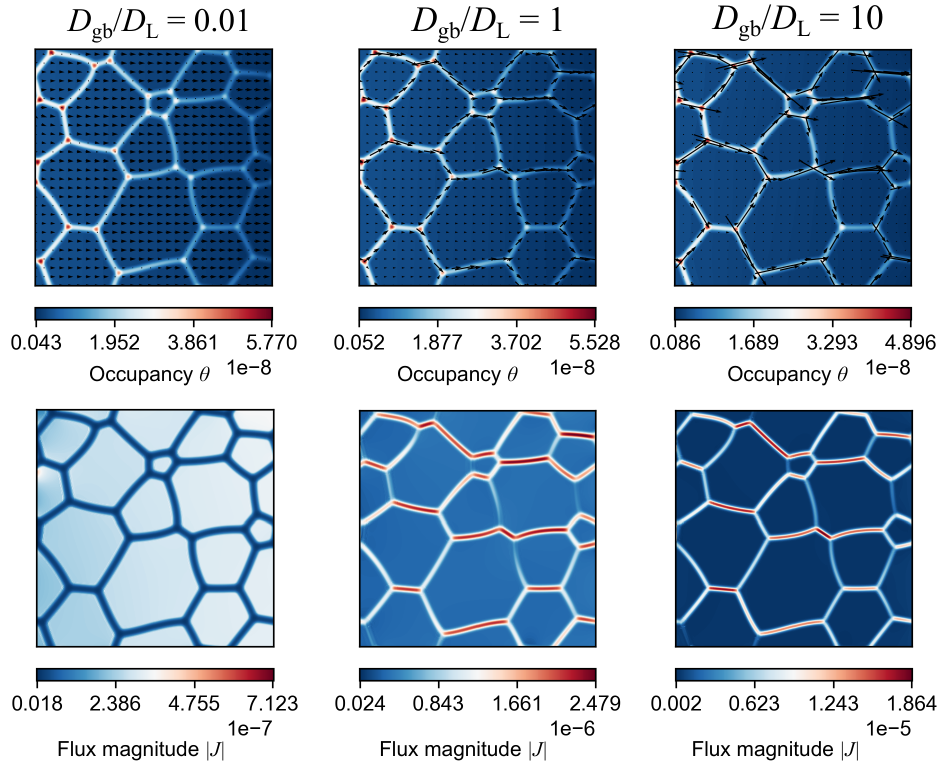}
  \centering
  \caption{Results for selected values of $D_\mathrm{gb}/D_\mathrm{L}$. The first row shows the occupancy field superimposed by the flux vector field $\vec{J}$. The second row shows the corresponding flux magnitude field $|J|$}
  \label{fig: permeation_flux}
\end{figure*}

In order to understand the interaction of GB trapping and diffusion at equilibrium, we analyze the full-field results of the flux in Fig.(\ref{fig: permeation_flux}). The upper row of Fig.(\ref{fig: permeation_flux}) shows the occupancy fields for three cases $D_\mathrm{gb}/D_\mathrm{L} =$ 0.01, 1 and 10 superimposed by the flux vector fields $\vec{J}$ and their magnitudes in the lower row. As discussed earlier, the occupancy fields are almost identical. In the case $D_\mathrm{gb}/D_\mathrm{L} =$ 0.01, the flux magnitude at the GBs is smaller than that in the lattice, while it is larger and increases with increasing $D_\mathrm{gb}$ for the cases $D_\mathrm{gb}/D_\mathrm{L} =$ 1 and 10. Furthermore, $\vec{J}$ is directed along the GBs, where the GBs aligned in the direction of the external flux have larger magnitude and it decreases with increasing the angle with the external flux down to perpendicular GBs. Additionally, the interconnectivity of GBs with small angles lead to higher flux magnitude. It was shown that for columnar nano-grained sputtered tungsten, hydrogen flux increased with increasing the grain's aspect ratio, i.e. GBs are more elongated along the direction of the external flux \cite{DiazRodriguez2022}. Van den Eeckhout \cite{Eeckhout2017} showed that cold rolled samples with elongated grains restored large percentage of the pre-deformation flux after annealing heat treatment. This resulting microstructure did not undergo recrystalization while the dislocations were annihilated, which was the reason for restoring the flux. We believe that the elongated GBs also played a role in restoring the flux.

\begin{figure*}[hbt!]
  \includegraphics[width=15cm]{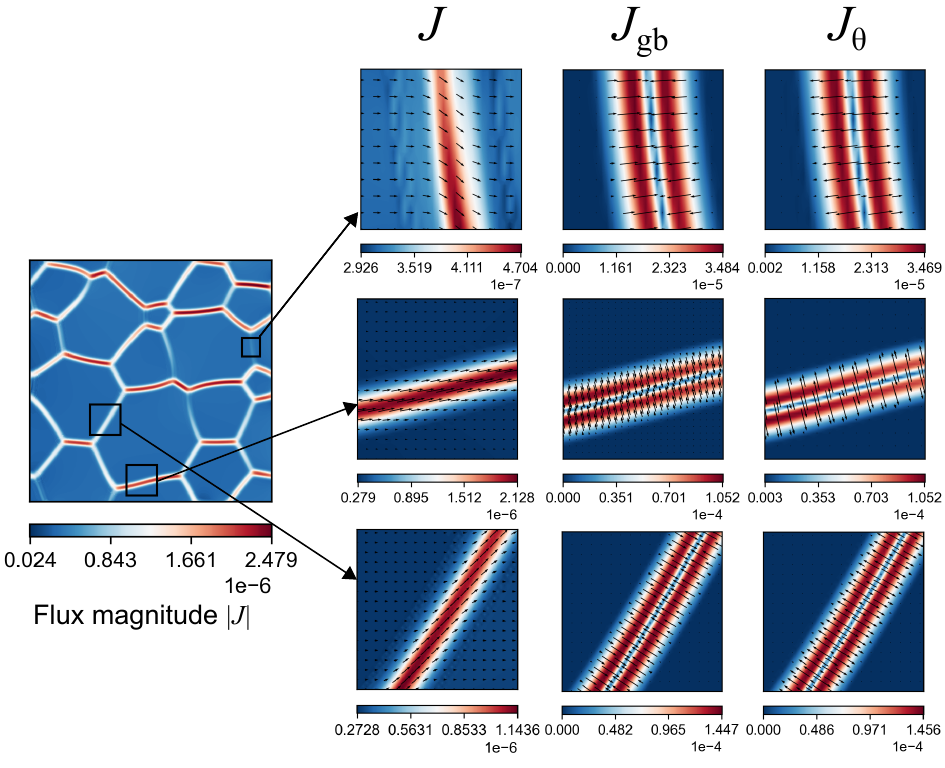}
  \centering
  \caption{The flux field for $D_\mathrm{gb}/D_\mathrm{L}=1$ and three cases of GB alignment with respect to the direction of the external flux field. The magnified figures show the total flux field $J$ and its constituent GB driven flux  $\vec{J}_\mathrm{gb}$ and concentration gradient driven flux $\vec{J}_\mathrm{\theta}$.}
  \label{fig: gb_con_flux}
\end{figure*}

In order to understand the flux behavior at the GBs, Fig.(\ref{fig: gb_con_flux}) shows the flux vector field for $D_\mathrm{gb}/D_\mathrm{L} =$ 1 for three cases of GB alignment with respect to the external flux. The small insets show a magnification of these alignments for the total flux field $\vec{J}$ and its constituent trapping $\vec{J}_\mathrm{gb}$ and diffusion $\vec{J}_\mathrm{\theta}$ fluxes. For all the three alignments, it could be seen that the magnitude of $\vec{J}$ decreases with increasing the angle of the GB with respect to the external flux. $\vec{J}_\mathrm{gb}$ is always directed towards the center of the GB and its magnitude is maximum at the GB sides and decreases towards the center. This is because the gradient (slope) of the GB potential $g(\phi)$ decreases towards the center. $\vec{J}_\mathrm{\theta}$ on the other side is in direction opposite to the occupancy gradient and has two components. A large component which is equilibrating $\vec{J}_\mathrm{gb}$, and thus points away of the GB center. 

\begin{figure}[hbt!]
  \includegraphics[width=8cm]{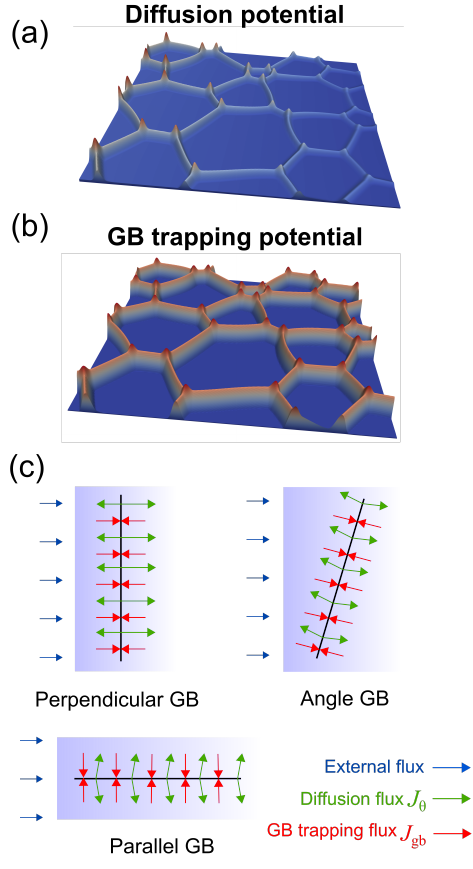}
  \centering
  \caption{3D surface of (a) the diffusion potential (occupancy field) at the steady state of the permeation simulation and (b) the GB trapping potential ($g(\phi)$). (c) A schematic showing how the alignment of the GB with respect to external flux field affect the direction of $\vec{J_\mathrm{\theta}}$ and $\vec{J}_\mathrm{gb}$ near the GB center.}
  \label{fig: flux_mechanism}
\end{figure}

The second component of $\vec{J}_\mathrm{\theta}$ is orders of magnitude smaller, and thus does not appear in Fig.(\ref{fig: gb_con_flux}). Since trapping increases the occupancy of the GBs, the occupancy gradient along GBs is larger as can be seen from the 3D surface of the occupancy in Fig.(\ref{fig: flux_mechanism}a). This component is not opposed by $\vec{J}_\mathrm{gb}$ since its potential $g(\phi)$ does not have a gradient along the external flux as shown in Fig.(\ref{fig: flux_mechanism}b). The resultant $\vec{J}$ along the GB is due to this small component of $\vec{J}_\mathrm{\theta}$ and is thus orders of magnitude smaller as shown in Fig.(\ref{fig: gb_con_flux}). Additionally, the magnitude of the small component of $\vec{J_\mathrm{\theta}}$, and thus $\vec{J}$ along the GB, depends on the angle of the GB with respect to the external flux as shown schematically in Fig.(\ref{fig: flux_mechanism}c). For perpendicular GBs, the occupancy gradient is also perpendicular to the GB, and thus, $\vec{J_\mathrm{\theta}}$ gets quickly opposed by $\vec{J}_\mathrm{gb}$. For parallel GBs, the component of $\vec{J}_\mathrm{\theta}$ has no opposing flux from $\vec{J}_\mathrm{gb}$ and is therefore maximum. Following the same reasoning, GBs with an in-between angle represent the intermediate case. In addition to the aforementioned discussion, the magnitudes of these interacting fluxes is directly proportional to both $D_\mathrm{gb}$ and $\Delta E_\mathrm{gb}$.

Therefore, trapping does not necessarily mean hydrogen can not move in the GBs, but rather its movement is confined within the GBs. In the case of uptake simulation, trapping makes GBs act like sinks for hydrogen and the enrichment is determined by $\Delta E_\mathrm{gb}$. Steady-state is reached when an equilibrium is established between $\vec{J}_\mathrm{\theta}$ and $\vec{J}_\mathrm{gb}$. In the case of permeation experiments, there will always be an external concentration gradient that will result in an external flux. In this case, the steady-state will be reached when an equilibrium is established between $\vec{J}_\mathrm{\theta}$ and both $\vec{J}_\mathrm{gb}$ as well as the external flux. The flux of hydrogen in the GBs will depend on the interaction between trapping and diffusion. The GB enrichment due to trapping will result in higher concentration gradients compared to the lattice. The interplay between $\Delta E_\mathrm{gb}$ and $D_\mathrm{gb}$ will determine the role of GBs whether they are hydrogen sinks or high diffusion paths. Having discussed the steady-state, next we discuss the early stages of permeation.

\begin{figure*}[hbt!]
  \includegraphics[width=13cm]{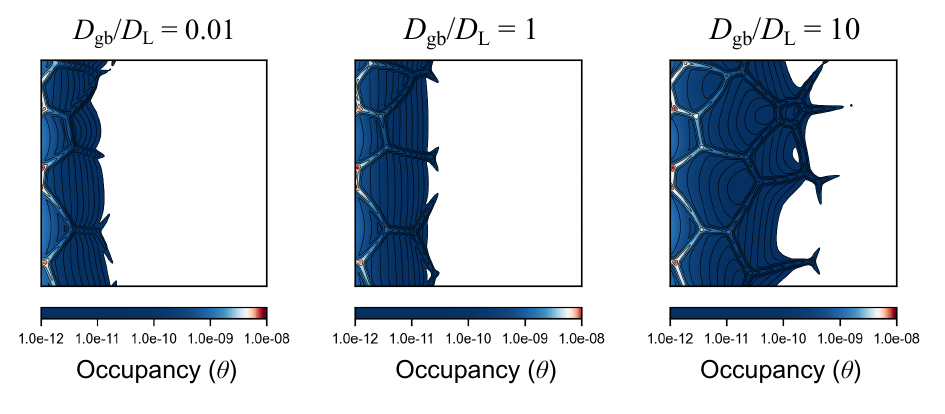}
  \centering
  \caption{A snapshot at early stage ($7.8 \times 10^{-3} \mathrm{s}$) of permeation simulation showing the effect of $D_\mathrm{gb}/D_\mathrm{L}$ on the advancement of occupancy field front. A log-scale is used to better represent the iso-contour lines.}
  \label{fig: Permeation_early}
\end{figure*}

The occupancy fields at early permeation time $7.8 \times 10^{-3}$ s for $D_\mathrm{gb}/D_\mathrm{L}=$ 0.01, 1 and 1 are shown in Fig.(\ref{fig: Permeation_early}). The occupancy fields are shown on log-scale with a cut-off value of $10^{-12}$ to identify the field advancing front and iso-contour lines. As expected, the advancing front is faster with increasing $D_\mathrm{gb}$. Nevertheless, the ratio $D_\mathrm{gb}/D_\mathrm{L}$ affects the shape of the advancing front. First, in all the three cases, the GBs are always ahead of the lattice. Second, a pattern could be seen for the shape of the advancing front. When $D_\mathrm{gb}<D_\mathrm{L}$, the front has a convex shape at the lattice part since it has higher diffusivity. However, due to trapping, GBs absorb hydrogen from the nearby lattice, making them always ahead of the lattice even when $D_\mathrm{gb}<D_\mathrm{L}$. The advancing front is almost a straight line in the lattice for $D_\mathrm{gb}<D_\mathrm{L}$. In this case, the high concentration gradient along the GBs increases their flux making them ahead of the lattice. , for $D_\mathrm{gb}>D_\mathrm{L}$, the advancing front is convex in the lattice. The GBs are ahead due to both their high diffusivity as well as the trapping high concentration gradient. The effect of $\Delta E_\mathrm{gb}$ on the GB flux will be discussed in the next section. 

\subsection{Effect of grain boundary trap-binding energy on hydrogen permeation}
\label{section: PermeationEnergy}

\begin{figure*}[hbt!]
  \includegraphics[width=16cm]{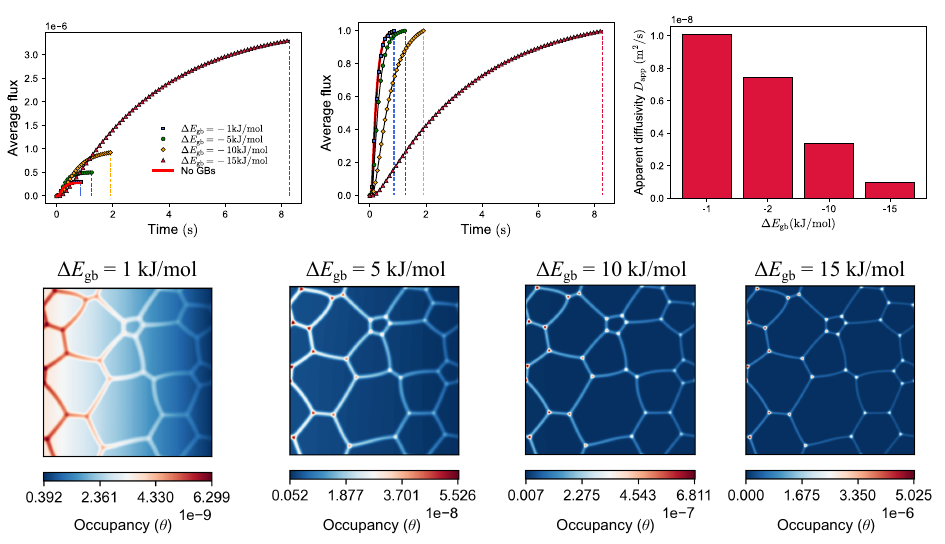}
  \centering
  \caption{The effect of GB trapping and diffusivity on hydrogen occupancy and flux in the early stages of the permeation test.}
  \label{fig: Permeation_Hseg}
\end{figure*}

The effect of $\Delta E_\mathrm{gb}$ on the evolution of the average and normalized flux as well as $D_\mathrm{app}$ are shown in the upper row of Fig.(\ref{fig: Permeation_Hseg}) and the occupancy fields in the lower row. Increasing $\Delta E_\mathrm{gb}$ increases the steady-state flux as well as the time to reach steady-state. Increasing $\Delta E_\mathrm{gb}$ also increases the GB occupancy. As described in the previous section, this leads to a higher occupancy gradient in the presence of external flux, which leads to increased diffusion flux along the GBs. On the other side, increasing $\Delta E_\mathrm{gb}$ leads to increased time to reach the equilibrium occupancy values in the GBs, and thus the time to reach steady-state. Again, $D_\mathrm{app}$ reflects this delay in reaching steady-state with increasing $\Delta E_\mathrm{gb}$, however, not the flux increase. Therefore, using $D_\mathrm{app}$ for studying hydrogen trapping requires further consideration since it only qualitatively represents the effect of $\Delta E_\mathrm{gb}$ on the time to reach steady-state, but not the steady-state flux. 

\begin{figure*}[hbt!]
  \includegraphics[width=16cm]{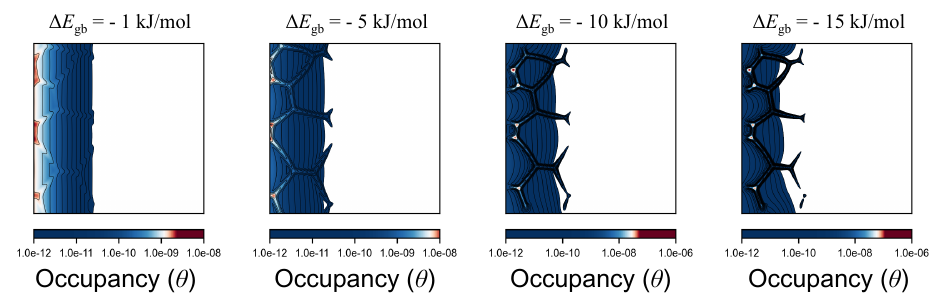}
  \centering
  \caption{A snapshot at early stage ($7.8 \times 10^{-3} \mathrm{s}$) of permeation simulation showing the effect of $\Delta E_\mathrm{gb}$ on the advancement of occupancy field front. A log-scale is used to better represent the iso-contour lines.}
  \label{fig: Permeation_early_Hseg}
\end{figure*}

The effect of $\Delta E_\mathrm{gb}$ on the early permeation time $7.8 \times 10^{-3} \mathrm{s}$ is shown in Fig.(\ref{fig: Permeation_early_Hseg}). It could be seen that for all values of $\Delta E_\mathrm{gb}$, the occupancy advancing front is almost the same and is at the grain boundaries, however, the difference is in the lattice advancement. For the low trapping energy $\Delta E_\mathrm{gb} = -1 \mathrm{kJ/mol}$, the difference between the lattice and GB advancing fronts is very small and is almost a straight line. With increasing $\Delta E_\mathrm{gb}$, the lattice diffusion becomes more delayed compared to the GBs, and its advancing front becomes more convex. This is a characteristic difference when the GBs advancing front is ahead of the lattice is due to higher trapping or due to higher GB diffusivity. For the first case, the lattice front is convex, while it is concave for the latter. As described in the previous section, GB trapping makes GBs to act like sinks that absorb hydrogen from the nearby lattice, depleting them and making them more convex. This effect increases with increasing $\Delta E_\mathrm{gb}$.

\section{Conclusions}

We performed a full-field computational study on hydrogen diffusion and GB trapping. A phase-field based RVE was used, which can explicitly model the effect of GBs, triple and higher-order junctions. Consequently, two novel phase-field compatible formulations for hydrogen equilibrium distribution as well as diffusion and trapping were derived. The latter was used to analyze the interaction of hydrogen diffusion and trapping, where the GB trap-binding energy $\Delta E_\mathrm{gb}$ and diffusivity $D_\mathrm{gb}$ were used as parameters. Two sets of simulations were performed: hydrogen uptake and permeation. The main outputs of the study could be summarized as follows:

\begin{itemize}
  \item The independent variable in the mass transport equation is the unitless total occupancy $\theta$, which could be directly compared to the thermodynamic equilibrium. The partitioning of the total occupancy into lattice and GB occupancy could be done in post-processing using the GB function $g(\phi)$. Furthermore, the conversion to units of $\mathrm{mol/m^3}$ could be done using the lattice and GB molar volumes $V_\mathrm{m}$ and $V_\mathrm{gb}$ respectively. 
  
  \item The total flux equation is composed of two interacting fluxes, the conventional Fickian diffusion flux, which is in the direction opposite to the occupancy gradient. The second is a trapping flux that is always directed towards the GB center. The factors determining the interaction of these fluxes include the boundary conditions, $\Delta E_\mathrm{gb}$, $D_\mathrm{gb}$ and the orientation of the GBs with respect to the direction of the external flux. 
  
  \item Increasing $\Delta E_\mathrm{gb}$ increased the GB occupancy and the RVE averaged hydrogen uptake. It was also shown that the maximum GB occupancy is a function of the local lattice occupancy, which is a direct manifestation of Oriani's local equilibrium hypothesis. 
  
  \item While $D_\mathrm{gb}$ directly affected the GB and RVE averaged flux, it had a negligible effect on the hydrogen uptake. Surprisingly, increasing $\Delta E_\mathrm{gb}$ also increased the GB and RVE averaged flux. This was attributed to the increased occupancy gradient along GBs due to trapping. 
\end{itemize}

\section*{Acknowledgement}

AH is supported by Research Foundation -- Flanders Marie Skłodowska-Curie actions - Seal of Excellence (MSCA SoE FWO), grant number 12ZZN23N. BK acknowledges the National Research Foundation of Korea (NRF) by grant NRF-2018R1A6A1A0302552617 of the Priority Research Program under the Ministry of Education, and in part by grant NRF-2021R1A2C1004540 under the Ministry of Science and ICT. AH would like to thank Prof Ingo Stienbach for the valuable discussions about the multiphase-field method and the choice of segregation potential. 

\FloatBarrier

\bibliographystyle{unsrtnat}
\bibliography{Hydrogen_GB}

\end{document}